

\input epsf

\rightline {SU-ITP-94-14}
\rightline {hep-th/9405103}
\rightline {May 1994}
\title{Comment On  A Proposal By Strominger
\foot{Work supported in part by NSF grant PHY89-17438 }}
\author{Leonard Susskind \foot{email:
susskind@dormouse.stanford.edu}}
\address {Deparment of Physics, Stanford University, Stanford, CA
94305-4060}

\abstract

Strominger has proposed an interesting concrete realization of
Hawking's
idea that information is lost in black hole evaporation. In this note
we
demonstrate that a straightforward interpretation of Strominger's
model
leads to a complete breakdown of the conditions for using statistics
for
analyzing the results of experiments. The probabilities produced by
the
theory are operationally meaningless.
\bigskip
\endpage

\Ref\Hawking{S.~W.~Hawking \journal Phys. Rev. & D14 (76) 2460.}

\Ref\strom{A.~Strominger, {\it Black Holes and Open Strings,}
preprint UCSBTH-94-13, May 1994, hep-th/9405094.}

\Ref\peski{T.~Banks, L.~Susskind , and M.~E.~Peskin \journal Nucl.
Phys. & B244 (84) 125.}

Consider the proposition that the probability for an experiment
denoted
$A$ to lead to a given result $B$ is $P(A;B)$. This is a meaningless
assertion unless the following two conditions are met.

1)  The experiment can be performed many times. In particular the
initial
conditions of the experiment are reproducible. In order to keep the
experiments independent they should ideally be performed in widely
separated regions of space time.

2)  The results of the individual experiments must be statistically
independent. This implies that the probability for the outcomes of an
ensemble of $N$ identically  prepared events  to be $B_1, B_2,
B_3,\ldots B_N$ must be

$$
P(A;B_1 \ldots B_N  )= P(A;B_1) P(A;B_2) \ldots P(A;B_N)
\eqn\fact
$$

This requirement can be considered to be a consistency condition on a
theory in the following sense. If we regard the ensemble of
experiments to
itself be a single experiment, the theory can be directly applied to
it. Using
the same rules of calculation that were used for the single event A,
we should
directly confirm eq. \fact\ .  In ordinary quantum field theory these
requirements are insured by the cluster decomposition property.

In recent years the validity of the usual operational rules of
quantum
mechanics has been questioned principally in the context of black
hole
formation and evaporation. According to Hawking [\Hawking ] the
process in
question inevitably leads to a loss of quantum coherence so that an
initial
pure state of matter will evolve to a mixed density matrix. It is of
obvious
interest to ask whether specific versions of Hawking's proposal
satisfy the
consistency condition of eq. \fact\ . Strominger [\strom ] has put
forward one such proposal which is specific enough to analyze from
this viewpoint. we shall see that a straightforward interpretation of
this proposal fails the test.

Strominger does not consider 3+1 dimensional gravity. His
construction is
based on the more easily analyzable 1+1 dimensional  models of Callen
Giddings Harvey and Strominger. It is argued that one particular
version of
the CGHS model is isomorphic to an open string theory. The physical
space-
time is identified with the world sheet of a semi-infinite string.
The target
space of string theory is the space of fields that propagate on the
world sheet. According to the model, black hole formation and
evaporation is always
accompanied by the splitting off of a piece of space which is
represented by
an open string. The emitted open string carries off the information
that fell into the black hole.

Before considering the implications of Strominger's model one thing
needs
to be clarified. In this paper the terms, experiments and their
outcome, refer to operations carried out in the usual way by
observers living in the universe.
We are not interested in the rules for ``superobservers'' who perform
experiments involving ensembles of universes. We are interested in a
set of
rules to be used by future experimental physicists like ourselves.

Consider an event in which matter in an initial state  $A$ creates  a
black
hole which subsequently evaporates producing an open string in state
$i$
and final state $B$ of the semi-infinite universe. The amplitude for
this
transition is

$$
S(A|B, i)
\eqn\amp
$$

The final state of the semi-infinite universe will not generally be
pure.
According to the most straightforward rule it will be described by
the density
matrix

$$
\sum_i S^*(A|B,i) S(A|B',i)
\eqn\den
$$

Eq.[\den ]   can be pictorially represented as in Fig (1) in which
the two sides
represent $S$ and $S^*$
\vskip 15pt
\vbox{{\centerline{\epsfsize=3.0in \hskip 2cm \epsfbox{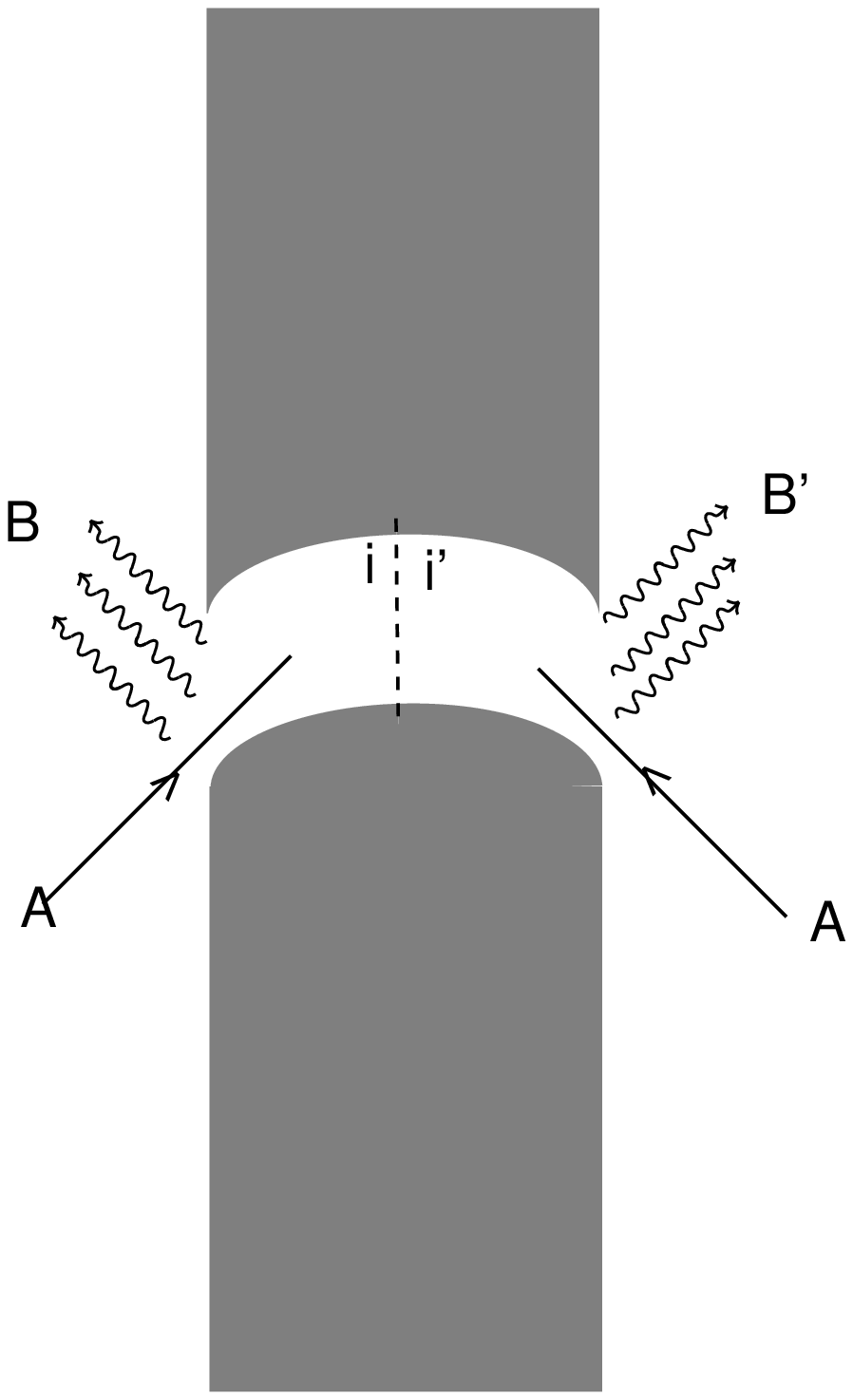}}}
\vskip 12pt
{\noindent{\tenrm FIGURE 1. The density matrix. }}
{\noindent{\tenrm The white region represents the world sheet. The
shaded region is not part of the universe.}}
\vskip 15pt}

An experiment to determine if information is lost can not be carried
out with a single event. We are therefore led to consider an event in
which the initial conditions  $A$ are replicated many times at widely
separated times. For simplicity consider just two copies of the
incident state. The event is illustrated in Fig(2).
\vskip 15pt
\vbox{{\centerline{\epsfsize=3.0in \hskip 2cm \epsfbox{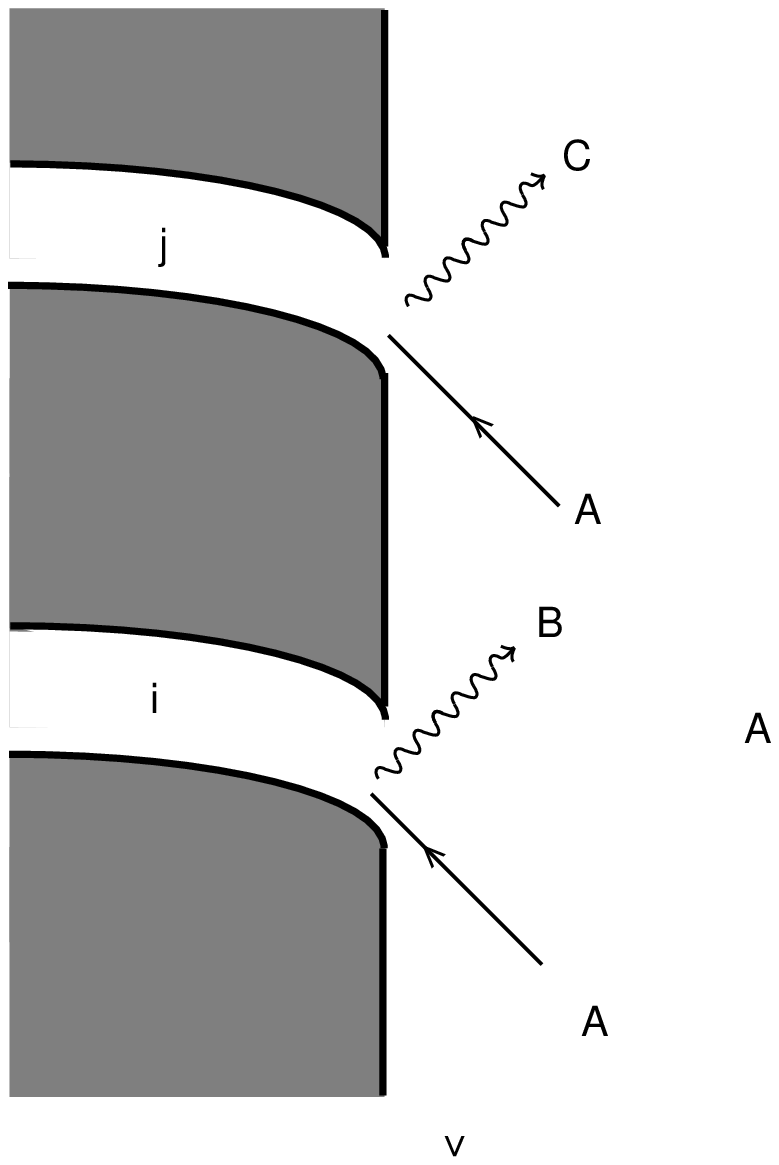}}}
\vskip 12pt
{\centerline{\tenrm FIGURE 2. The density matrix for two identical
events.}}
\vskip 15pt}

The two emitted open strings are emitted at widely
separated world-sheet times but in identical target space
configurations. In
string theory they would be considered as identical particles and
subject to
the rules of either Fermi- Dirac or Bose-Einstein statistics. The
implication is that in squaring the amplitude and summing over the
open strings, two
distinct ways of making the identification must be made. This is
shown in Fig
(3a) and (3b).

\vskip 15pt
\vbox{{\centerline{\epsfsize=3.0in \hskip 2cm \epsfbox{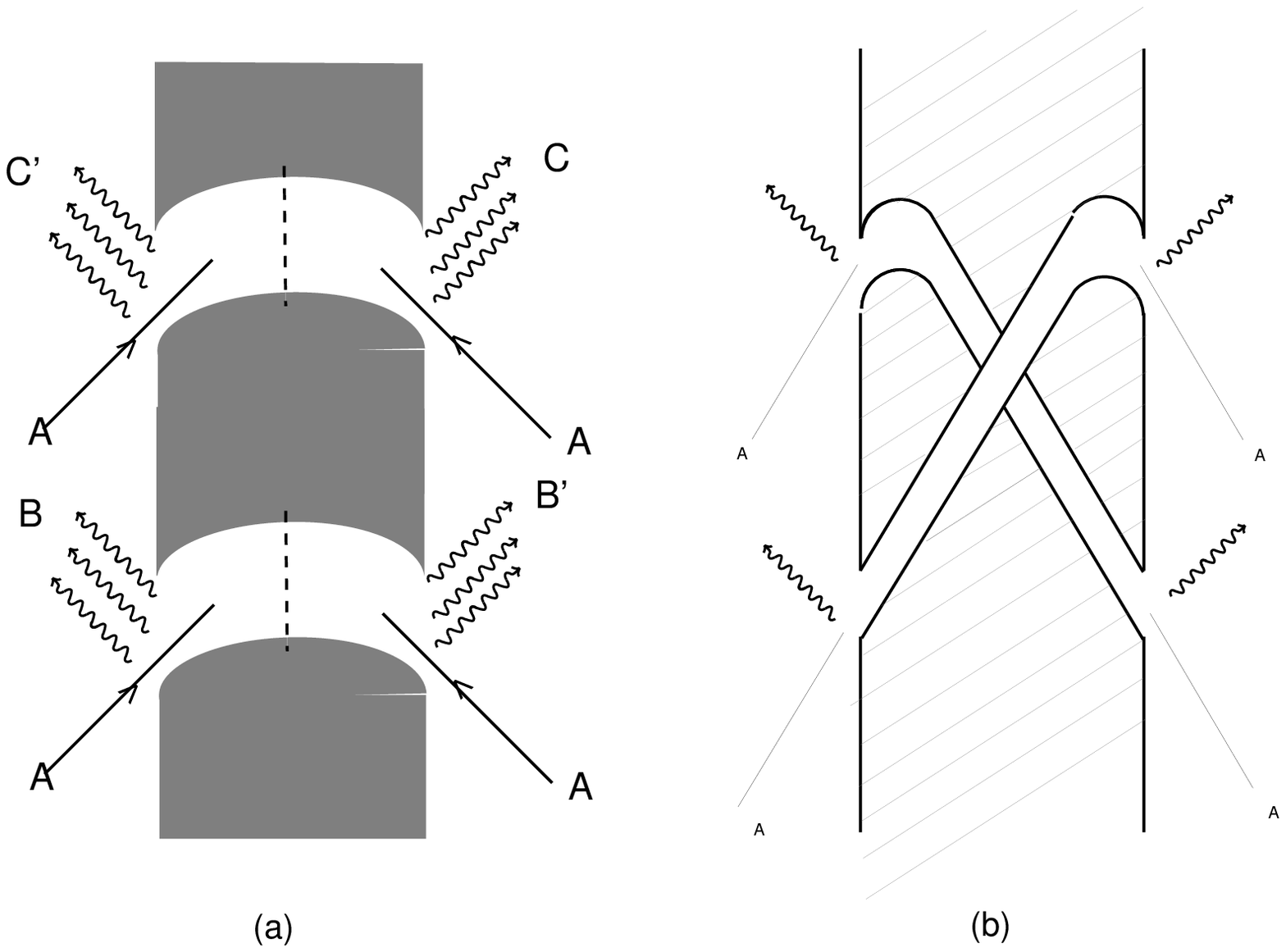}}}
\vskip 12pt
{\noindent{\tenrm FIGURE 3a. First way of identifying open strings.
}}

{\noindent{\tenrm FIGURE 3b. Second way of identifying open strings;
in this figure the
cross-hatched region is not part of the worldsheet. }}
\vskip 25pt}

Thus the final density matrix is

$$
\eqalign{&S(A|B,i) S^*(A|B',i) S(A|C,j) S^*(A|C',j) \cr
&\pm S(A|B,i) S^*(A|C',i) S(A|C,j) S^*(A|B',j) \cr }
\eqn\dblden
$$
(The plus and minus refer to the target space statistics)

The first term in eq.[\dblden] is all that would be expected if the
events were
statistically independent. As an example of the peculiar features of
eq. [\dblden]
suppose in both experiments we post select the same final state B. In
the
theory of a single event, the probability is

$$
\sum_i S(A|B,i)S^*(A|B,i) = P_A (B)
\eqn\sing
$$

For the double event the probability obtained from eq.[\dblden] is

$$
P_A(B,B) = P_A(B)^2\pm P_A(B)^2
\eqn\dbl
$$

For example if the open strings are target space fermions the double
event
can not occur at all while for bosons the probability is double the
value
required by eq.[\sing] .  The point is not that we have encountered
some unusual behavior of probabilities but that the very conditions
for the use of statistics have broken down. We emphasize again that
unless experiments can be performed repeatedly, under identical
circumstances with statistically independent results there is no
meaning to the probability of a given outcome.

It is interesting to see what happens if we try to repair the damage
by
making a rule which would only allow identification of strings if
they are
emitted at similar world sheet times. In other words suppose the open
strings carry some memory of when they are emitted and can only be
treated as identical if emitted at nearby times. In this case there
would be a tight correlation between the times that the branching
processes take place on the two sides of the figures representing the
evolution of the density matrix. This leads to exactly the situation
discussed by Banks, Peskin and Susskind [\peski ] and therefore to
non conservation of energy.

It is possible that the rules of string theory may have a less
straightforward interpretation, perhaps along the lines of Coleman's
wormhole calculus. How such a reinterpretation would translate into a
set of
rules useful to ordinary observers is not clear.

Acknowledgenents:

The author thanks Andy Strominger for a preliminary copy of Ref
[\strom]
\refout
\vskip 2cm

\vfill\eject
\end